# Magnetic shielding simulation for particle detection

**Sara R. Cabo**[1,3,4,a] , **Sergio Luis Suárez Gómez**[2,3,b], **Laura Bonavera**[1,3,c], **Maria Luisa Sanchez**[1,3,d], **Jesús Daniel Santos**[1,3,e], **Francisco Javier de Cos**[3,4,f]

[1] Departamento de Física, Universidad de Oviedo, Calvo Sotelo 18, 33007 Oviedo, Asturias, Spain
[2] Departamento de Matemáticas, Universidad de Oviedo, Calvo Sotelo 18, 33007 Oviedo, Asturias, Spain
[3] Instituto Universitario de Ciencias y Tecnologías Espaciales de Asturias (ICTEA), Independencia 13, 33004 Oviedo, Asturias, Spain
[4] Departamento de Explotación y prospección de Minas, Universidad de Oviedo, Independencia 13, 33004 Oviedo, Asturias, Spain

**Abstract** Cherenkov-type particle detectors or scintillators use as a fundamental element photomultiplier tubes, whose efficiency decreases when subjected to the Earth's magnetic field. This work develops a geomagnetic field compensation system based on coils for large scale cylindrical detectors. The effect of different parameters such as the size of the detector, the distance between coils or the magnetic field strength on the compensation using a basic coil system composed of circular and rectangular coils is studied. The addition of coils of very specific geometry and position to the basic configuration is proposed in order to address the compensation in the areas of the detector where it is more difficult to influence, in order to minimize the loss of efficiency. With such improvement, in the considered simulated system, more than 99.5% of the photomultiplier tubes in the detector experience an efficiency loss of less than 1% due to the effect of the magnetic fields.

## 1 Introduction

The current research in the field of particle physics, namely the study of fundamental particles and forces that constitute matter and radiation, sometimes requires the use of indirect detection methods. This is the case of the detection of neutrinos and cosmic-rays that might be produced when dark matter candidates annihilate or decay [1].

Some of the most commonly used instruments for such measurements are large scale Cherenkov detectors and scintillators [2–4]. In a Cherenkov detector, a very large volume of dielectric material such as ultrapure water is used. When the particles under consideration interact with the molecules of the dielectric material, the charged particles resulting from this interaction move through the medium with a phase velocity higher than that of light, producing so-called Cherenkov light in the visible or ultraviolet range. This light is produced in the form of a cone in the propagation direction of the particle and later detected as rings [5].

On the other hand, in detectors based on scintillator material, the passage of ionizing particles interacting with the molecules of the material causes excitation of the atoms. These atoms are rapidly de-excited, giving rise to light [6].

In the specific type of detector considered in the present manuscript, the light produced is collected by photodetectors, which cover the entire inner surface of the detector. From the amount of light collected and its characteristics, the energy value, direction and type of the incident particle are derived. In general, photomultiplier tubes (PMTs) are used because of their fast response, high gain, and high signal-to-noise ratio.

In order to obtain the most accurate results possible, it is essential to maximize the efficiency of the photodetectors. In particular, such efficiency decreases when they are subjected to a magnetic field because it affects the secondary electrons produced in the PMTs. Even a very weak magnetic field (even lower than $10^2$ mG) might affect the efficiency and thus the whole experiment. An important source of external magnetic field that has to be taken into account when planning such kind of experiments on Earth is the geomagnetic field. The Earth's magnetic field is therefore a cause of efficiency loss that must be avoided [7].

[a] e-mail: rodriguezcsara@uniovi.es (corresponding author)

[b] e-mail: suarezsergio@uniovi.es

[c] e-mail: bonaveralaura@uniovi.es

[d] e-mail: mlsr@uniovi.es

[e] e-mail: jdsantos@uniovi.es

[f] e-mail: fjcos@uniovi.es

## 1.1 Geomagnetic field

The geomagnetic field or Earth's magnetic field that affects in this particular case the performance of the detector's photomultipliers has its origin in the Earth's outer core. The fundamentals of its generation are currently explained by the dynamo theory [8].

The Earth's outer core is composed of highly conductive molten iron and nickel. The propagation of heat from the solid inner core due to some processes such as the release of energy by compression, generates convection currents in the fluids that compose the outer core. The movement of the conductive fluids produces electrical currents which generate a magnetic field. In turn, this varying magnetic field again creates electric current, producing a dynamo effect that self-sustains the Earth's magnetic field. This effect can be expressed by the so-called magnetic induction equation.

$$\frac{\partial B}{\partial t} = \nabla \times (u \times B) + \eta \nabla^2 B \tag{1}$$

where $u$ is the velocity of the fluid, which in this case is just the molten metal of the earth's outer core, $B$ is the magnetic field and $\eta$ is called magnetic diffusivity and is just the inverse of the product of the electrical conductivity and the magnetic permeability of the medium.

The Coriolis effect produced by the planet's own rotation on the charged particles moving in the outer core causes them to acquire spiral trajectories, which means that the field lines are aligned with the axis joining the north and south poles.

Currently, the Earth's magnetic field behaves like a dipole and forms an angle of 15° with the planet's rotational axis. Its magnitude varies over the Earth's surface between approximately 250 mG and 650 mG, being more intense at the poles and less so at the magnetic equator, although its intensity and orientation changes over time, small variations can occur with a frequency of only milliseconds to secular variations, i.e. on time scales of more than a year, generally reflecting changes in the Earth's metallic core, including reversal of the magnetic north and south poles [9].

With the aim of removing this effect, two of the most used methods that are being applied in several experiments with these type of detectors are the establishment of magnetic shields on the photodetectors, as in the case of the BOREXINO neutrino detector [10], and the design of a geomagnetic field compensation system based on coils. The latter approach has been successfully used in experiments such as Super-Kamiokande [11] or MiniCLEAN [12] and is being considered for detectors still under construction such as JUNO [13] and Hyper-Kamiokande [14]. Moreover, the coil's method is also of great importance in occasions where the active material is ultrapure water, as is the case for detectors such as Super-Kamiokande or JUNO, as the metals normally used as magnetic shields oxidize and therefore deteriorate in contact with water [15].

The aim of this article is to study through simulations an optimum compensation system for the geomagnetic field. In particular, our study focuses on cylindrical geometry detectors since they are among the most commonly used, and we aim at studying the effect on the system of parameters that play fundamental roles in the design of this type of experiments such as the dimensions of the detector, the distance between the photodetectors and the inner walls of the tank and the distance between coils. The presence of detectors in diverse locations around the planet, at different latitudes and therefore subject to different values of the geomagnetic field [16], also makes it necessary to study the compensation of the geomagnetic field according to the location, i.e. the value of the total field strength as well as its components.

The first section establishes the characteristics of the simulated detector, the compensation system and the physical principles on which the subsequent simulations are based. In the following section, the geomagnetic field compensation under several parameters such as, the detector size, the height to width ratio, the distance between coils, the distance of the photodetectors to the walls and the intensity of the geomagnetic field are studied, in order to find the optimal coil system for the compensation, according to the detector measurements. The addition of further coils, beyond the basic configuration, are proposed in very specific positions to address areas of the detector where geomagnetic field compensation is more complicated and thus minimize the loss of efficiency of all photodetectors in the detector. The last section presents our conclusions.

## 2 Objective, design of the experiment, design of the system and simulation

### 2.1 Objective

External magnetic fields are one of the causes of the decrease in photon collection efficiency of PMTs. This is because low-energy electrons travel along a long path in a vacuum and their trajectories are affected by even a slight magnetic field such as the magnetic field produced by our own planet, causing an anode sensitivity variation. This loss of efficiency is particularly marked when the direction of the external magnetic field is perpendicular to the photomultiplier axis. On the other hand, if the direction is parallel to the photomultiplier axis, the effect of magnetic field is negligible [17].

A magnetic field of less than 50 mG does not cause a loss of efficiency. From this value onwards, the efficiency starts to decrease progressively, reaching a loss of 1% with a magnetic field of 100 mG in the direction perpendicular to the PMT axis, which increases rapidly with the gradual increase of the magnetic field [13, 14, 17]. It is then established as an objective a value of the magnetic field perpendicular to the axis of the PMTs of, at most, 100 mG for the maximum possible number of PMTs to ensure correct operation

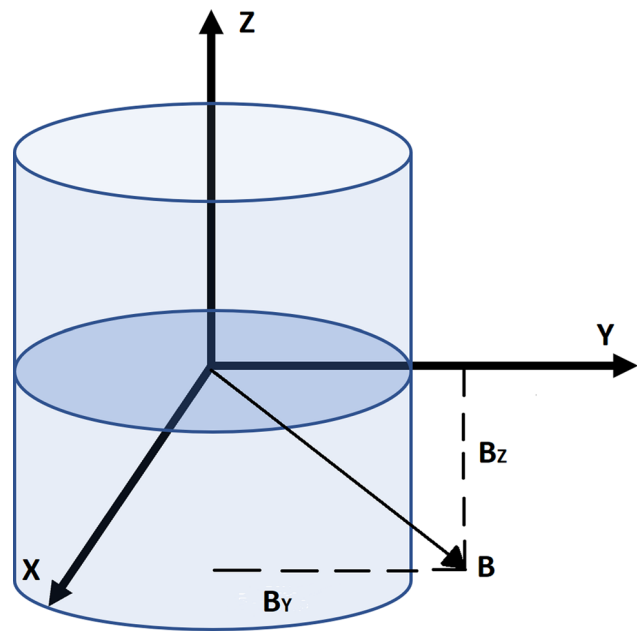

**Fig. 1** Schematic representation of the generic cylindrical detector under consideration. The reference system used is also shown, with origin at the centre of the detector, as well as an example of a geomagnetic field vector

of the detector. To achieve this goal, the approach of this paper will be to design a coil system, placed on the inner surface of the detector, that compensates the geomagnetic field perpendicular to the PMTs, so that the residual field is less than the threshold value of 100 mG.

The objective of this paper is to find an effective compensation system that guarantees a good functioning of PMTs in a generic cylindrical detector. It is also a primary goal to analyse several geometrical parameters in the experiment, that influence the choice of coil characteristics.

2.2 Design of the system

The detector scheme is laid out as a cylinder, at the centre of which is the origin of coordinates of the Cartesian reference frame. The geomagnetic field can be represented at any point by a three-dimensional vector. The reference system is shown in Fig. 1. In order to achieve simplicity in the design of the compensation system, the reference system can be rotated so that the $B_X$ component of the Earth's magnetic field is always zero. By identifying the Z-axis with the direction of the axis of the cylindrical detector, as shown in Fig. 1, it is always possible to rotate the coordinate system about the Z-axis so that the geomagnetic field vector is contained in the YZ plane and, consequently, the $B_X$ component cancels out. This means that only two coil systems need to be introduced, one to compensate the geomagnetic field in the Z-axis direction and one to compensate it in the Y-axis direction.

The basic system proposed in this article for the compensation of the geomagnetic field inside the detector consists therefore of two sets of coils that are shown in red in Figs. 2 and 3. For visual purposes this configuration of coils is split into two figures even if they constitute a single system. One set of circular coils is placed horizontally to compensate for the $B_Z$ component of the geomagnetic field (see Fig. 2) and a second set of vertical rectangular coils centred on the Y axis is added to the system to compensate for the magnetic field $B_Y$ introduced in that direction (see Fig. 3). It is important to note that both sets of coils will create a magnetic field with a component in the X-direction that is not meant to compensate any counterpart produced by the geomagnetic field, meaning that we are introducing a slight excess in that direction. However, the value of this component is expected to be an order of magnitude smaller than the value of the field created by the coils in the Y and Z directions, small enough not to contribute significantly to the magnetic field perpendicular to the PMTs to be minimized.

The coils are located right on the inside wall of the tank. The circular coils have the same radius as the detector, and the rectangular coils are circumscribed. Their height is the same as that of the detector, and the length of their horizontal sides varies according to the position. Thus, when talking about the distance between the PMTs and the walls, it is equivalent to the distance between PMTs and coils, hence this parameter has an influence on the geomagnetic field compensation.

To proceed with the optimization for the maximum efficiency of the PMTs, the magnetic field has to be calculated at the exact position of each PMT. Therefore, these positions have to be determined in the simulation. In this case, PMTs of 50 cm in diameter have been considered, taking as reference the most current models of photomultipliers designed specifically for the study of high-energy physics, such as the Hamamatsu R12860, or the Hamamatsu R3600 model, used in the Super-Kamiokande neutrino detector [14]. The centre-to-centre distance between adjacent PMTs is set at 70 cm. The PMTs are installed on scaffolding inside the detector and define a second inner cylinder, so the effective detection volume is determined by the distance between the PMTs and the inner surface of the tank. The greater the distance, the smaller the detection volume. The free space between adjoining PMTs is considered to be covered by opaque black sheets to absorb incident radiation [11].

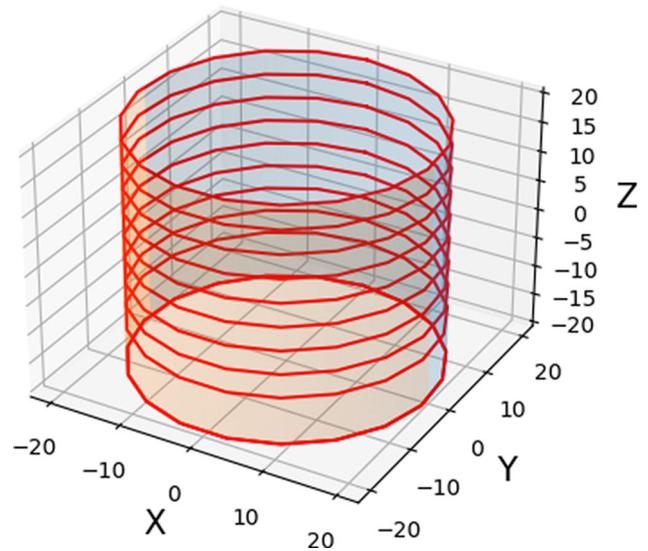

**Fig. 2** Simulation of circular coil system in red centred on Z-axis for geomagnetic field compensation in a tank of 40 m height and diameter

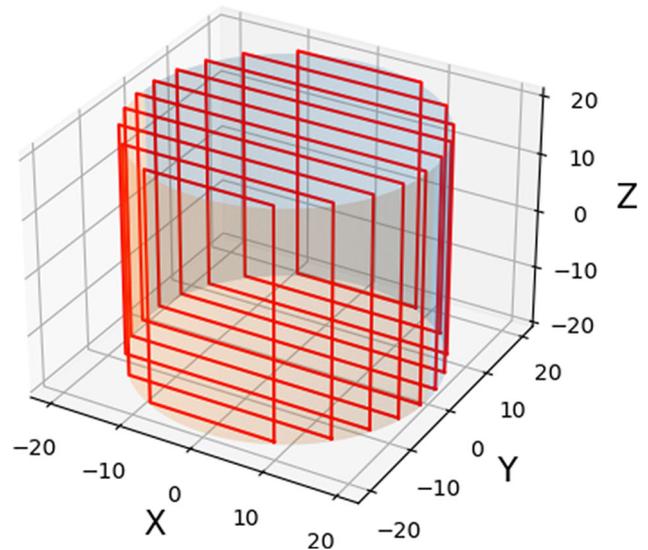

**Fig. 3** Simulation of rectangular coil system in red centred on Y-axis for geomagnetic field compensation in a tank of 40 m height and diameter

The current flowing through the coils generates, due to the Joule effect, a heat emission inside the detector which may contribute to the temperature rise of the water and, consequently, to the production of dark currents inside the photodetectors and, even, the proliferation of bacteria in the water, which reduces its transparency to Cherenkov photons [11]. The distribution of this heat in the water can also create convection currents that distribute the radon emitted by the PMTs and other electronic devices throughout the detector. Radon is one of the main sources of background radiation [14, 18]. The expression for the dispersed heat is as follows.

$$Q = I^2 R = I^2 \rho \frac{L}{S} \tag{2}$$

where I is the current flowing through the cable, R is the resistance, $\rho$ is the resistivity of the conductive material composing the cable, L is its length and S is its cross-section. Although detectors of this type usually incorporate a system for purifying and regulating the temperature of the water [11, 14] it is nevertheless advisable to choose a cable for the coils with the lowest possible resistance or, in other words, the largest cross-section in order to minimize the amount of heat dispersed.

### 2.3 Calculation of perpendicular magnetic field

The Earth magnetic field to be compensated is considered constant throughout the detector, as variations from one position to another are negligible in the absence of magnetic materials in the environment. By applying a certain value and direction of current intensity to the coils of the compensation system, a magnetic field opposite to the Earth's magnetic field is created. The sum of the two

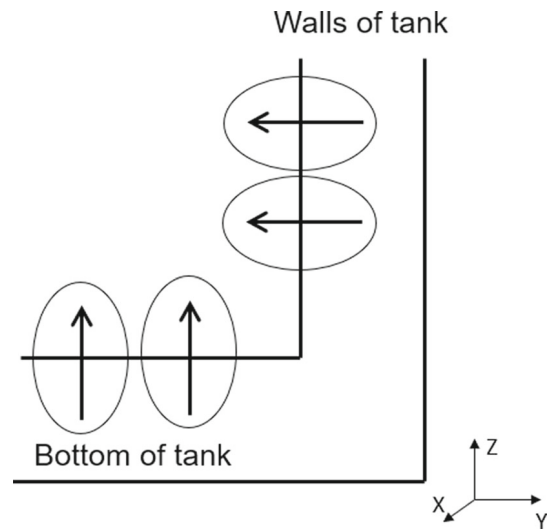

**Fig. 4** Orientation of PMTs in a cylindrical detector. The orientation depends on whether they are on the lids or on the walls. Ovals represent the PMTs and the arrows their orientations. The XYZ system of reference is also depicted

magnetic fields results in a residual magnetic field denoted as $\Delta B = (B_X, \Delta B_Y, \Delta B_Z)$. On the other hand, the residual magnetic field perpendicular to the PMTs axis is denoted as $\Delta B_{perp}$. This is the magnetic field whose modulus is to be minimized.

The direction of the magnetic field perpendicular to the axis of the photomultiplier tubes depends on their orientation and position in the detector. Figure 4 shows how PMTs are oriented according to their position in a cylindrical detector. The ovals represent the PMTs and the arrows their orientations; the solid lines are the walls of the tank of the Cherenkov detector, vertical walls and bottom of the tank. The XYZ system of reference is also depicted in the lower right corner.

The expression that determines the magnetic field perpendicular to its axis is different depending on whether the PMTs are on the walls or whether they are located in the upper or lower lids. The PMTs on the lids have the particularity that their axis is oriented in the direction of the Z-axis. This means that the $\Delta B_Z$ component of the residual magnetic field does not cause efficiency losses, but $B_X$ and $\Delta B_Y$ components do, as they are orthogonal to the Z-axis. The expression of $\Delta B_{perp}$ corresponding to these PMTs, easily calculated by means of both components, is shown in Eq. (3).

$$\Delta B_{perp}^{lids} = \sqrt{B_X^2 + \Delta B_Y^2} \qquad (3)$$

The case of wall PMTs is slightly more complex, as all three components of the residual magnetic field $B_X$, $\Delta B_Y$ and $\Delta B_Z$ contribute to the perpendicular magnetic field, since in this case, the photomultiplier axis is not parallel to any of the three coordinate axes. The expression for the corresponding perpendicular magnetic field is shown in Eq. (4).

$$\Delta B_{perp}^{walls} = \sqrt{(B_X \cdot sin\theta + \Delta B_Y \cdot cos\theta)^2 + \Delta B_Z^2} \qquad (4)$$

where $\theta$ is the angle with the $Y$ axis. To calculate the magnetic field created by each coil at the exact position of each photomultiplier, the law of Biot and Savart is used in the context of Maxwell's classical theory of electromagnetism. It has been calculated for circular and rectangular coils [19].

2.4 Design of the experiment

Several parameters have been identified that influence the compensation of the geomagnetic field by the proposed coil system. These are: the dimensions of the detector, the current intensity circulating through the coils, the distance between the PMTs and the internal surface of the detector, the distance between coils and the intensity of the geomagnetic field and its components.

In order to study the greater or lesser influence of the above variables, different simulations have been carried out, so that it is possible to identify the optimal values of such parameters that lead to a better compensation of the geomagnetic field.

Firstly, the optimal current intensity value to circulate through the different sets of coils in order to obtain the best results is studied. This optimum value varies depending on the distance between the coils, the distance between the PMTs and the walls and the size of the detector, so different simulations are carried out varying these parameters in an appropriate range.

An interesting parameter to take into account in order to evaluate the good functioning of the compensation system is the amount of PMTs subjected to a magnetic field above a certain threshold. The proportion of PMTs above 100 mG is evaluated for each detector model (i.e, different sizes of detector), and varying other geometrical parameters like distance between coils and distance to the walls, using the optimal intensity of current.

The next parameter to be studied is the geomagnetic field strength. To do this, the simulation of a detector of certain dimensions is repeated, varying the distance between coils and to the walls. In this case, different values of the total geomagnetic field strength are considered, assuming that all its components have the same value in order to appreciate only the influence of the total strength on the compensation. Subsequently, the proportion of PMTs with excess magnetic field is plotted as a function of latitude, fixing the value of the rest of the parameters, including the total geomagnetic field strength, which is considered to have an intermediate value. The reason for this study is that, given a value of the total geomagnetic field strength, the value of its components is also a factor influencing the compensation, which vary as a function of latitude.

The study of the influence of the different parameters is completed by simulating the geomagnetic field compensation in detectors with different height-to-diameter ratios. The optimization of the results is not only influenced by whether the detector is larger or smaller, but also by the value of this ratio, which is a significant factor to be taken into account.

Finally, it is illustrated by means of a concrete example how an optimal choice of parameters based on the results obtained in the previous studies allows to considerably reduce the proportion of PMTs subjected to excess magnetic field. Furthermore, additional measures, such as the addition of extra coils of specific geometry and position, are proposed to minimize the proportion of PMTs and thus reduce the efficiency loss to the fullest extent.

## 2.5 Simulation

The program used to simulate the geomagnetic field compensation in the detector consists of three Python modules. The advantage of the first module consists on the possibility of considering coils with any shape, not only the rectangular and circular ones, which have an analytical solution, but also any other shape with no analytical solution. In particular, circular, rectangular and elliptical coils are used. In addition, solenoids with the mentioned geometries are also defined. Another module aims to calculate the magnetic field created by the coils defined in the previous module in any coordinates indicated by applying directly Biot and Savart's law.

In the main module, where the two previous modules are imported, the characteristic parameters of the detector, such as its dimensions, are defined and the exact position of the photomultipliers inside the detector is established. Two functions are distinguished within the module. The central function establishes the position, geometry, size and current intensity of the coils. It calculates the total magnetic field, the sum of the geomagnetic field and the magnetic field created by all the coils, by components and the magnetic field perpendicular to the axis of each photomultiplier. The values returned by the function are the average perpendicular magnetic field over the entire detector and the percentage of photomultipliers that exceed the established upper limit of 100 mG of perpendicular magnetic field.

The second function built into the module constructs a histogram of the perpendicular magnetic field and a graphical representation of the cylindrical tank showing the position of all those photomultipliers exceeding 100 mG of residual perpendicular magnetic field.

Finally, a third function has been added to this module that implements a genetic algorithm to help optimize different parameters. A genetic algorithm is a programming technique used for the optimization of non-linear multivariate problems inspired by genetic and natural selection processes [20]. The starting point is a population of $N$ individuals, each of which is called a chromosome. Each chromosome is made up of different genes, which are nothing more than values of the parameters to be optimized. Thus, to optimize a number $n$ of parameters, each chromosome will contain $n$ values, one corresponding to each parameter.

The genes in this particular case will be the current intensity in the circular coils, the current intensity in the rectangular coils, the number of coils of each type and the distance between them.

On the other hand, the other key piece is the definition of a fitting function that takes these chromosomes as input and returns a measure of how good they are as solutions to the problem. In this case, the fitting function will be the program that has been used so far that calculates the residual perpendicular magnetic field over the different PMTs, so that it returns the number of photomultipliers subjected to less than 100 mG. The goal is then to maximize this number of PMTs.

By selecting the individuals that best fit the problem according to the fitness function, that is, those values of the genes which maximise the number of PMTs subjected to a perpendicular magnetic field of less than 100 mG, combining them to replace the worst-fitting individuals, and introducing random mutations to the genes during the various iterations of the programme, the algorithm should be able to return an optimal configuration of the system. Naturally, the greater the number of genes or parameters to be optimized and the greater the range of possible values for each, the greater the number of iterations required to reach convergence [21].

## 3 Results

The compensation of the geomagnetic field depends not only on the particular value of the geomagnetic field, but also on other parameters. In fact, the compensation is highly dependent on the dimensions of the detector and its components. In particular, the size of the detector, i.e. the value of its height and diameter, the distance between the coils and the distance between the position of the PMTs and the walls, are parameters that play an important role in the compensation of the geomagnetic field and therefore have to be taken into account when designing the detector.

To study the effect of the parameters listed above, each of them is varied within a range of values consistent with the current design of large scale detectors. In a first simulation, the cylindrical detectors that are being studied have the ratio between diameter and height set to one and cylinders of three different sizes are considered in order to cover as wide a spectrum as possible.

**Fig. 5** Proportion of PMTs with excess only with the set of rectangular coils as a function of intensity of current for different values of the distance between coils in a detector of 40 m of height and diameter and a distance from PMTs to the walls of 2 m

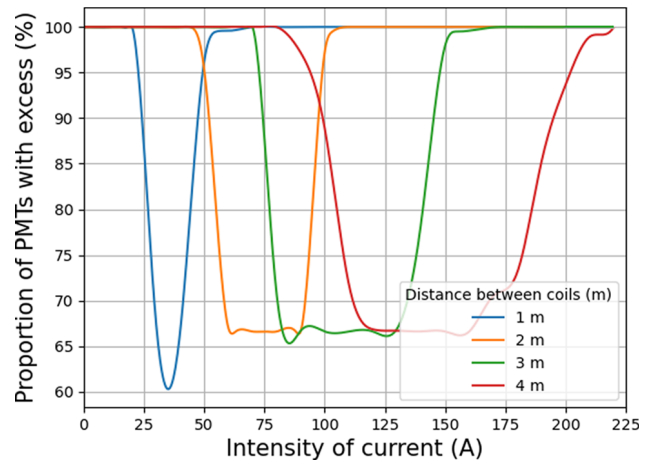

**Fig. 6** Optimal intensity of current for the set of rectangular coils as a function of distance between coils for different values of the distance to walls in a detector of 40 m of height and diameter

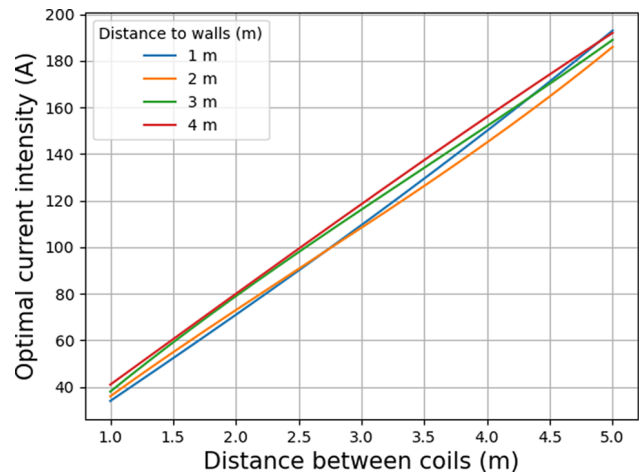

The distance between the coils varies between 1 and 5 m and the distance of the PMTs to the walls from 1 to 4 m. Since the PMTs are placed on supports, it is considered necessary to leave a space of at least 1 m between the PMTs and the walls specifically for such supports, which is why smaller distances between the PMTs and the walls are not simulated.

The magnitude of the Earth's magnetic field varies between 250 and 650 mG, reaching its maximum value at the magnetic poles and its minimum value at the magnetic equator [16]. For this first simulation, an intermediate magnetic field of 475 mG was chosen, although the effect on the compensation of fields with more extreme values are also checked with additional simulations.

### 3.1 Optimal current intensity

Once the value of the previously mentioned parameters (dimensions of the detector, distance between inner walls and PMTs and distance between coils) has been set, the proportion of PMTs subjected to excess residual perpendicular magnetic field, i.e. experiencing a magnetic field perpendicular to its axis greater than 100 mG, is minimized. For this purpose, the value of the current intensity applied to both sets of circular and rectangular coils is optimized. This optimal value of the current intensity is highly dependent on the distance between the coils, and less dependent on the size of the detector and the distance of the PMTs to the walls for a given value of the Earth's magnetic field. The greater the distance between the coils, the higher the current intensity to be applied to them as it is shown in Figs. 5 and 7. Alternatively, Figs. 6 and 8 show also this effect in addition to the influence of the distance from PMTs to the walls. The size of the detector considered was 40 m height and 40 m diameter, and the magnetic field was $B = 475$ mG, which corresponds to an average latitude of 45°.

As it can be seen in Figs. 5, 6, 7 and 8, the largest variation in current intensity occurs when the distance between the coils is changed. Increasing the distance between coils by 1 m results in an average increase of $\Delta I \approx 35\ A$ in the optimum current. The change in optimum current as a function of the distance from the PMTs to the walls is much less significant, although there is a gradual increase in current intensity as the distance to the wall increases for a fixed value of the distance between coils, which is more noticeable for the set of circular coils. However, in the case of 4 m and 5 m distance between coils, a decrease in the optimum current intensity is observed when the distance to the walls is increased from 1 to 2 m. It is also notable that, for fixed values of the distance between coils and to the wall, the optimum current intensity is always greater or equal for the set of rectangular coils compared to the set of circular ones.

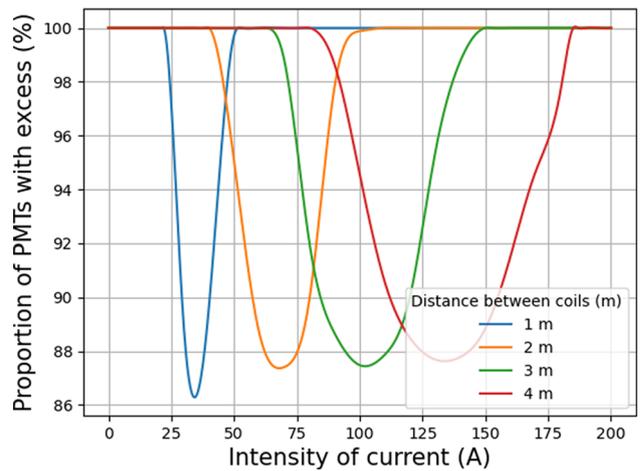

**Fig. 7** Proportion of PMTs with excess only with the set of circular coils as a function of intensity of current for different values of the distance between coils in a detector of 40 m of height and diameter and a distance from PMTs to the walls of 2 m

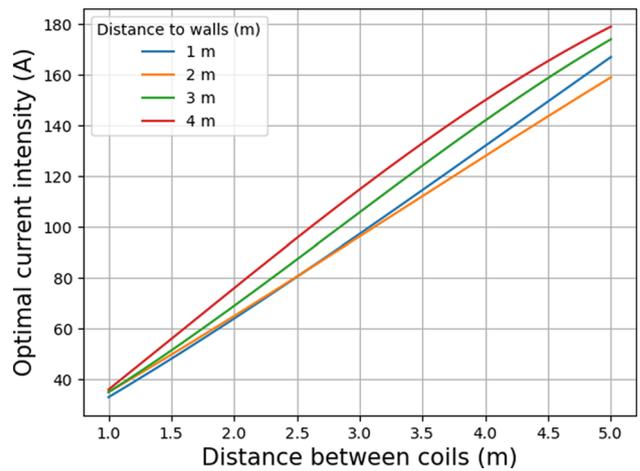

**Fig. 8** Optimal intensity of current for the set of circular coils as a function of distance between coils for different values of the distance to walls in a detector of 40 m of height and diameter

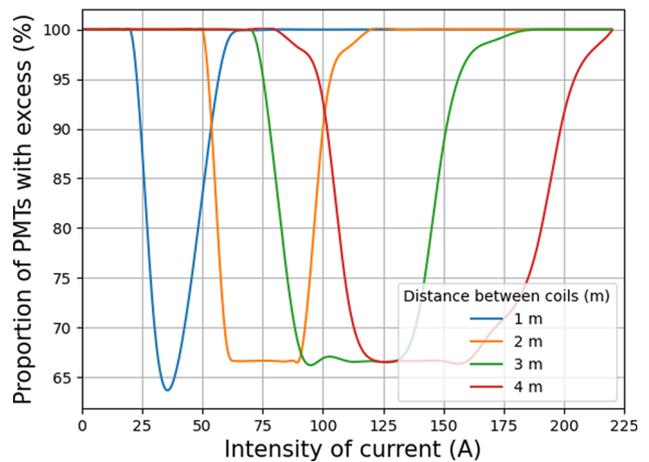

**Fig. 9** Proportion of PMTs with excess only with the set of rectangular coils as a function of intensity of current for different values of the distance between coils in a detector of 20 m of height and diameter and a distance from PMTs to the walls of 2 m

Figures 9, 10, 11 and 12 show the optimal currents of the rectangular and circular coil arrays considering the same value of the geomagnetic field, but this time in a detector of 20 m height and diameter, in order to compare the influence of the detector size on the variation of the optimal current intensities.

It is found that the range in which the optimum current intensity is situated for each value of the distance between coils is very similar to the previous case. The noticeable increase in current intensity with increasing distance between coils is the same, as is the average increase in current intensity per metre added between coils. It is also invariably observed that for a given value of the distance between coils and the distance to the walls, the optimum current for the set of rectangular coils is equal to or higher than the optimum current for the circular coils. In contrast to the larger 40 m detector, a smaller influence of the distance to the walls can be observed.

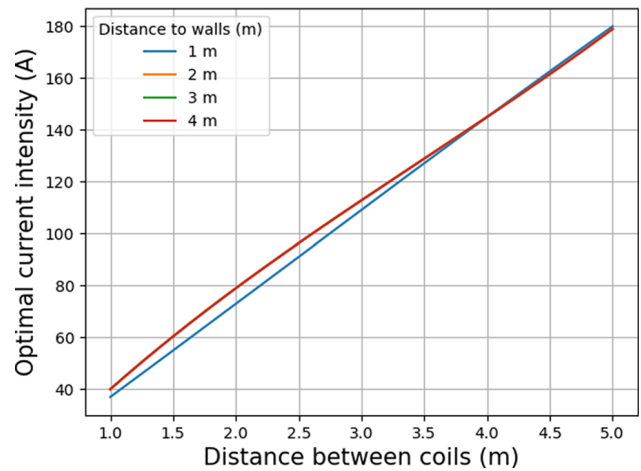

**Fig. 10** Optimal intensity of current for the set of rectangular coils as a function of distance between coils for different values of the distance to walls in a detector of 20 m of height and diameter

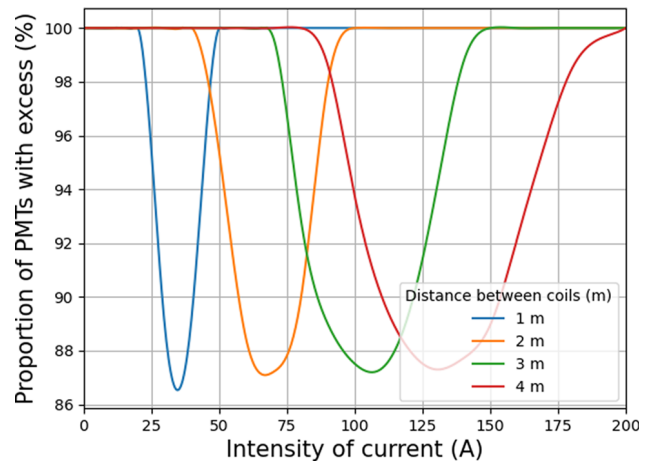

**Fig. 11** Proportion of PMTs with excess only with the set of circular coils as a function of intensity of current for different values of the distance between coils in a detector of 20 m of height and diameter and a distance from PMTs to the walls of 2 m

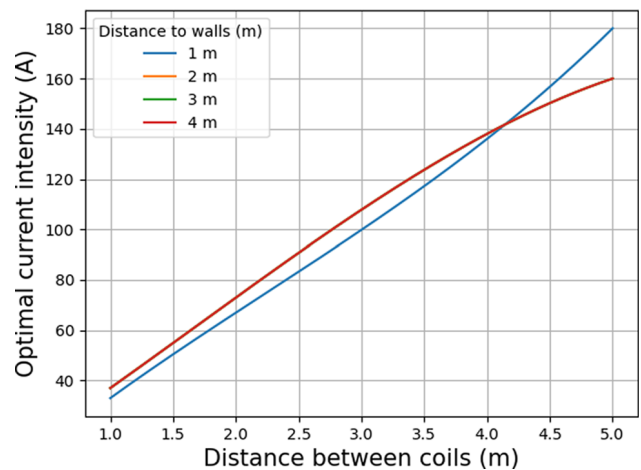

**Fig. 12** Optimal intensity of current for the set of circular coils as a function of distance between coils for different values of the distance to walls in a detector of 20 m of height and diameter

Given any fixed value of of the distance between coils, the general tendency in this smaller detector is for the optimum current intensity to stabilize with increasing distance of the PMTs from the walls. From a distance to the walls of 2 m, the value of the optimum current stabilizes, regardless of the distance between the coils, and remains at exactly the same value. It is for this reason that there appear to be only two functions represented, those corresponding to 3 m and 4 m distance from the walls take exactly the same values as the 2 m function. For instance, with a distance between coils of 4 m, the optimal value of current intensity is of 138

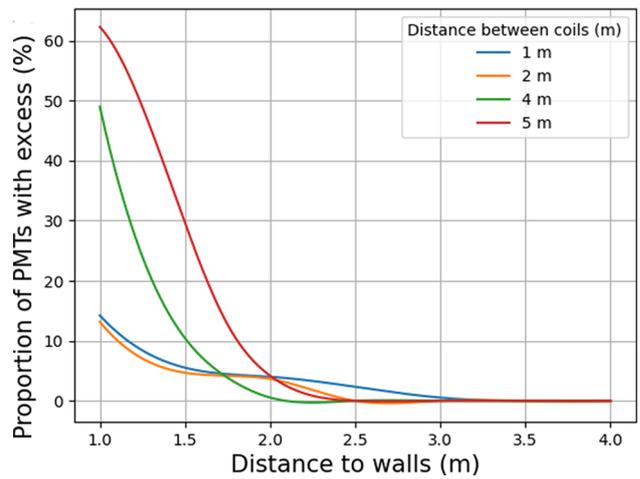

**Fig. 13** Proportion of PMTs with $\Delta B_{perp} > 100$ mG as a function of the distance of the PMTs to the loops for different loop spacings and a value of geomagnetic field of 475 mG in a detector with h = D = 20 m

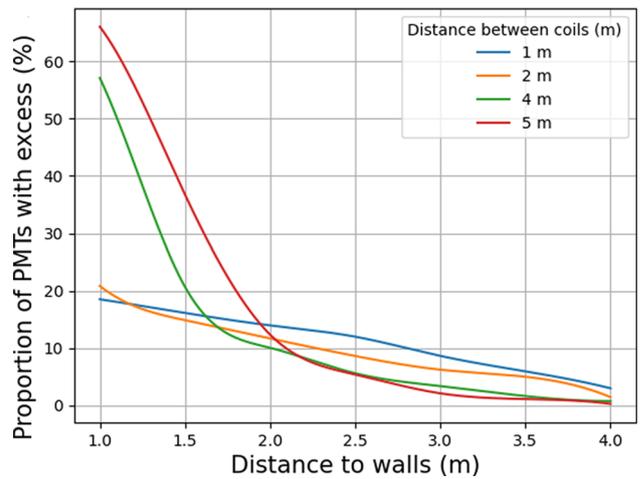

**Fig. 14** Proportion of PMTs with $\Delta B_{perp} > 100$ mG as a function of the distance of the PMTs to the loops for different loop spacings and a value of geomagnetic field of 475 mG in a detector with h = D = 40 m

A for a distance to walls equal to or greater than 2 m. The largest change corresponds to the distance to walls under 2 m. With a spacing of 5 m between coils, a decrease in the ideal current intensity is again observed when the distance to the wall is increased from 1 to 2 m.

3.2 Effects of size, distance to walls and between coils

The results of a first simulation are shown in Figs. 13, 14 and 15. Each of the plots shows the proportion of PMTs with excess residual perpendicular magnetic field as a function of the distance of the PMTs to the walls, where the coils are located, for different values of the distance between coils. Figure 13 shows the simulation for a detector with a height and diameter of 20 m, while Fig. 14 shows the simulation for a detector with a height and diameter of 40 m and Fig. 15 for a detector with a height and diameter of 60 m.

It can be noticed that the proportion of PMTs with excess of perpendicular residual magnetic field decreases as the distance of the PMTs to the walls increases. Given a certain value of the distance between coils, it is possible to decrease the proportion of PMTs with excess magnetic field to zero by simply separating the PMTs sufficiently from the walls, which would allow implementing a fairly simple compensation system based only on this presented pair of rectangular and circular coil arrays. However, it should be kept in mind that a larger distance between the PMTs and the inner surface of the detector implies a smaller photosensitive surface and a smaller volume of the material arranged for the interaction with the target particles, something to be taken into account especially for detectors of smaller dimensions, where an increase in this distance becomes more noticeable.

Another noteworthy aspect is the effect of the detector size on the proportion of PMTs with excess of magnetic field. The larger the detector, the higher this ratio is for a given distance between coils, denoted as $d_C$, and from PMTs to the walls, which will be denoted as $d_W$. Thus, while for a 20 m high and diameter detector it is sufficient to set the distance of the PMTs to the walls at $d_W = 2$ m to get the ratio to zero for a $d_C = 4$ m distance between coils, for a 60 m detector it is necessary to increase the distance to the walls to more than $d_W = 4$ m to achieve a zero ratio of PMTs with excess magnetic field for any distance between coils in the studied range. A distance from the PMTs to the walls of $d_W = 2$ m in a detector of 20 m implies a detection volume loss of $\Delta V = 3.07$ dam$^3$, which is 48.8% of the total tank volume. In contrast, a distance of $d_W = 4$ m at the detector of 60 m leads to a decrease of $\Delta V = 59.21$ dam$^3$ of effective volume. This is 34.9% of the total.

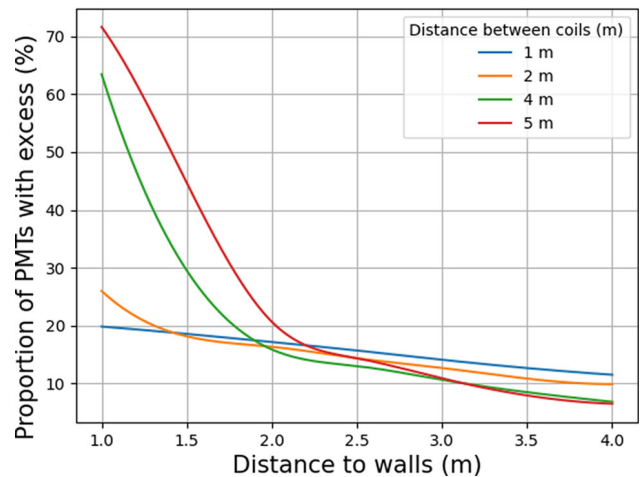

**Fig. 15** Proportion of PMTs with $\Delta B_{perp} > 100$ mG as a function of the distance of the PMTs to the loops for different loop spacings and a value of geomagnetic field of 475 mG in a detector with h = D = 60 m

The optimal spacing between coils varies with the dimensions. It varies significantly with the distance between the PMTs and the walls, and more slightly with the size of the detector. The greater the distance to the walls, the greater the optimal spacing.

For a detector of 20 m in height and diameter, the best distance between coils when the separation between the PMTs and the wall is between $d_W = 1$ m and $d_W = 1.75$ m is $d_C = 2$ m. Although distances to walls of less than $d_W = 1$ m are not considered, it can be extrapolated from Fig. 13 that in this range the best distance between the coils would be $d_C = 1$ m. From a distance of $d_W = 1.75$ m to the walls, the best distance between coils becomes $d_C = 4$ m, although a slight increase in the separation between PMTs and walls reduces the proportion of PMTs with excess magnetic field to zero for a distance between coils greater than $d_C = 2$ m.

For a detector of intermediate size, 40 m in height and diameter, the same type of behaviour is observed, with the difference that the intersections at which the optimal distance between coils changes occur for a larger distance between the PMTs and walls. In this case, the optimal distance between coils when the distance to the wall is $d_W < 1.25$ m is $d_C = 1$ m, while a distance between coils of $d_C = 2$ m is optimal if the distance to the walls is in the range $d_W \in [1.25, 1.85]$ m. For a wall distance $d_W \in [1.85, 2.15]$ m, the optimum spacing is $d_C = 4$ m and for $d_W > 2.15$ m, $d_C = 5$ m spacing produces better results.

If the detector size is 60 m, the behaviour remains the same and, again, it can be seen that the optimal spacing changes for a larger distance to the wall compared to the case of a 40 m detector. A distance between coils of $d_C = 1$ m is optimal until a distance between walls and PMTs of $d_W = 1.45$ m is reached. A distance between coils of $d_C = 2$ m is optimal in the range $d_W \in [1.45, 1.95]$ m, of $d_C = 4$ m is optimal for $d_W \in [1.95, 3.0]$ m. For $d_W > 3.0$ m, a $d_C = 5$ m spacing gives the best results, although in this case no spacing will achieve a zero proportion of PMTs with excess magnetic field. To achieve this with just this coil configuration in a detector of this size, the distance to the wall would have to be further increased. To minimize this increase, the distance between coils would also have to be increased, following the behaviour analysed so far.

The optimal distance between the coils depends strongly on the dimensions of the detector. This is because, in order to obtain the best compensation results, both circular and rectangular coils must be symmetrically placed between the two ends of the detector. Consequently, both the height and diameter values must be divisible by the distance between the coils. Figure 16 shows the proportion of remaining PMTs with excess magnetic field as a function of both circular and rectangular coil spacing for a detector with a height and diameter of 40 m, wall distance of $d_W = 2$ m and geomagnetic field value of $B = 475$ mG. In the previous simulations, discrete values of the distance between coils have been chosen. In this case, a continuous range of values is considered.

It can be observed that in the range considered the values of the distance between coils that offer the best results are those that divide the dimensions of the detector, that is, $d_W = 2$ m, $d_W = 2.5$ m, $d_W = 3.07$ m and $d_W = 4$ m. All of them lead to minima in the proportion of PMTs with excess magnetic field, with the case of $d_W = 4$ m being particularly noteworthy, which offers the best optimization, as deduced based on the results in Fig. 14.

When carrying out the previous simulation, that gives rise to the results in Figs. 13, 14 and 15, the same distance between coils has been considered for both rectangular and circular sets. However, it may well be that setting a different distance between coils for each of the sets leads to better results. Tables 1 and 2 show the results for a detector with height and diameter of 20 m and of 40 m, respectively. They list the proportion of PMTs with excess magnetic field for different values of the distance between rectangular and circular coils spanning from 1 to 5 m, optimizing the intensity of current, $d_W = 2$ m from the PMTs to the walls and a geomagnetic field of 475 mG with $B_Y = B_Z = 335.87$ mG components.

Given that in both cases the distance to the walls has been set at $d_W = 2$ m, based on the results shown in Figs. 13 and 14 it is to be expected that the case where there is $d_C = 4$ m between both rectangular and circular coils is the most favoured among those where the distance between coils is the same for both sets and, indeed, this is the case for both 20 m and 40 m detectors. It is noteworthy, however, that it is also the configuration that gives the best results, including those where the distance between coils in both sets is different. It can then be concluded that, if the detector design allows it, the most optimal configuration involves the same distance between coils for both sets and the choice of such distance is established as a function of the distance of the PMTs to the walls and the size of the detector.

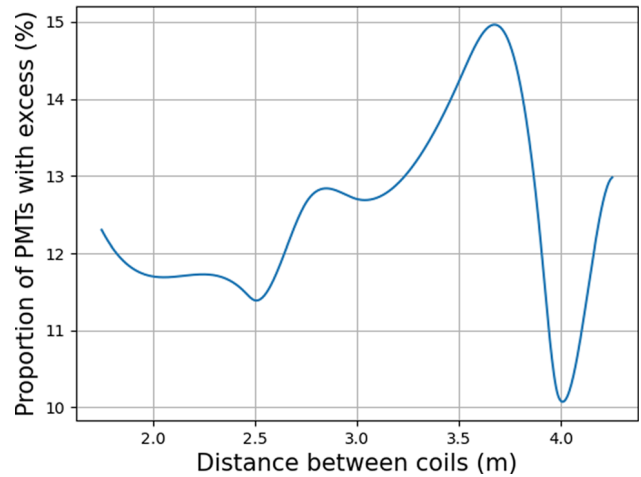

**Fig. 16** Proportion of PMTs with excess of magnetic field as a function of the distance between coils for a distance $d_W = 2$ m to walls, geomagnetic field of $B = 475$ mG and a detector of $h = D = 40$ m

**Table 1** Proportion of PMTs with excess of magnetic field for different distances between coils for rectangular and circular coils in a detector of $h = D = 20$ m (%)

| D. circular coils (m) | D. rectangular coils (m) | | | |
|---|---|---|---|---|
| | 1 | 2 | 4 | 5 |
| 1 | 3.02 | 4.68 | 5.56 | 9.62 |
| 2 | 4.60 | 3.65 | 1.70 | 8.71 |
| 4 | 4.39 | 1.95 | 0.55 | 7.05 |
| 5 | 6.09 | 5.26 | 3.77 | 4.19 |

**Table 2** Proportion of PMTs with excess of magnetic field for different distances between coils for rectangular and circular coils in a detector of $h = D = 40$ m (%)

| D. circular coils (m) | D. rectangular coils (m) | | | |
|---|---|---|---|---|
| | 1 | 2 | 4 | 5 |
| 1 | 13.99 | 14.59 | 16.11 | 19.19 |
| 2 | 14.77 | 11.7 | 14.44 | 17.03 |
| 4 | 14.16 | 10.94 | 10.08 | 16.42 |
| 5 | 14.39 | 10.25 | 12.85 | 12.38 |

If this choice is not possible due to the specific dimensions of the detector, the choice of the best basic configuration is more complex and depends on the size of the detector.

In the Table 1 corresponding to a detector of 20 m in height and diameter, it can be seen, as already mentioned, that the optimum configuration is the one that involves a distance between coils of $d_C = 4$ m for both sets. In case the distance between coils is required not to be the same for both sets, it is observed that, fixed the distance in one of the sets, the best results are obtained if the distance between coils of the other set is, again, $d_C = 4$ m. It should be noted, however, that there are two exceptions to this behaviour. For a distance $d_C = 1$ m in either set, the best results are obtained if the distance is $d_C = 1$ m also in the other. On the other hand, if the distance between rectangular coils is $d_C = 5$ m, the optimal distance between circular coils is also $d_C = 5$ m.

The results associated with a larger detector such as those shown in the Table 2 for a detector of 40 m in height and diameter are even more complex, as fewer general conclusions can be drawn. Again, the best results are obtained for a distance between turns $d_C = 4$ m for both rectangular and circular coils. However, in addition to this fact, if one considers setting a different coil distance for both sets, the previous apparent rule that the best results are achieved for, given a fixed coil distance for one of the sets, $d_C = 4$ m for the other no longer follows. It is notable, instead, that the minima in the ratio found for $d_C = 1$ m for circular and rectangular coils and $d_C = 5$ m between circular coils fixed $d_C = 5$ m for rectangular ones hold.

The same question can arise with respect to the distance between PMTs and walls. In the simulation associated with Figs. 13, 14 and 15 the same distance is considered between the PMTs on the lids and the base and top of the detector and the PMTs on the sides with the side surface of the detector. Figures 17 and 18 show for a detector of 20 m of height and diameter and a detector of

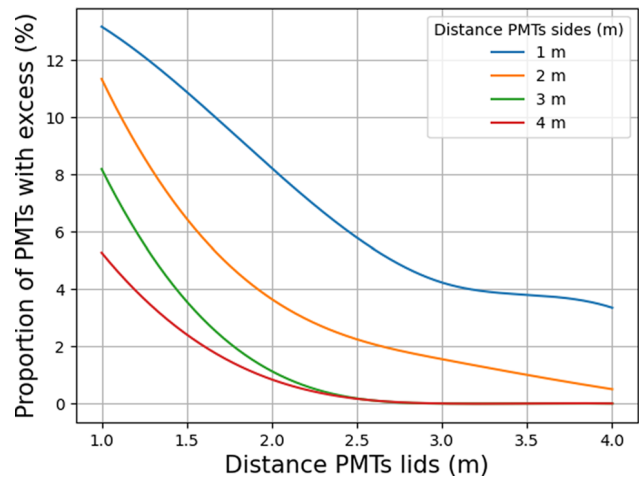

**Fig. 17** Proportion of PMTs with excess of magnetic field for different values of the distance to the side walls and the caps in a detector of $h = D = 20$ m

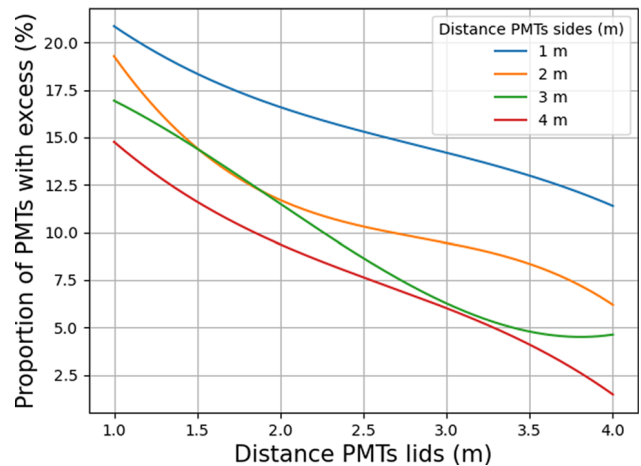

**Fig. 18** Proportion of PMTs with excess of magnetic field for different values of the distance to the side walls and the caps in a detector of $h = D = 40$ m

40 m of height and diameter respectively the ratio of PMTs with excess magnetic field for different values of the distance to the detector surface for the PMTs at the base and those at the walls, $d_C = 2$ m between coils and an Earth magnetic field of 475 mG with $B_Y = B_Z = 335.87$ mG components.

In this case, the behaviour of the system as a function of the variation of the distance to the detector surface distinguishing between distance between the PMTs from the bases to the caps and distance from the PMTs from the walls to the side surface is much simpler to analyse than in the previous study as a function of different distance between coils for the different assemblies. The relationship is much clearer and is as one would expect. For both the 20 m height and diameter detector and the 40 m detector, it is observed that the proportion of PMTs with excess perpendicular magnetic field decreases as either of the two distances to the tank surface increases. It can be concluded that the greater the distance separating the PMTs from the walls, the smaller the proportion of PMTs subjected to more than 100 mG of perpendicular magnetic field. It should be noted, again, that a larger distance implies a smaller interaction volume and photosensitive surface.

Figure 19 shows the decrease of the effective detection volume as the distance between the PMTs and the detector walls increases. The simulation has been carried out for the three detector sizes considered so far, with $h = D = 20$ m, $h = D = 40$ m and $h = D = 60$ m. For simplicity, the same distance between the top and bottom PMTs and the respective tank tops and between the side PMTs and the walls is considered.

The larger the dimensions of the detector, the greater the loss of detection volume per metre added of distance to the walls. For example, increasing the distance of the PMTs to the walls from 1 to 2 m on a 40 m detector results in a loss of detection volume of 6.45 dam$^3$, whereas on a 20 m detector this loss would be only 1.36 dam$^3$. Nevertheless, in the case of the 40 m detector, this decrease represents a loss of 14.97% of the initial detection volume. For the 20 m detector, this same proportion would rise to 29.69%. It is therefore important to find a balance between increasing the distance of the PMTs to the walls to optimize the magnetic field compensation and a photodetector surface large enough to achieve the particular objectives of the experiment.

Given the complexity of detecting the target particles of this type of detectors, such as neutrinos, maximising the detection volume is crucial. Although with the coil configuration proposed in this article it is possible to satisfactorily reduce the effect of the magnetic field on the PMTs in most cases, reducing the distance of the PMTs to the detector walls too much can overcomplicate

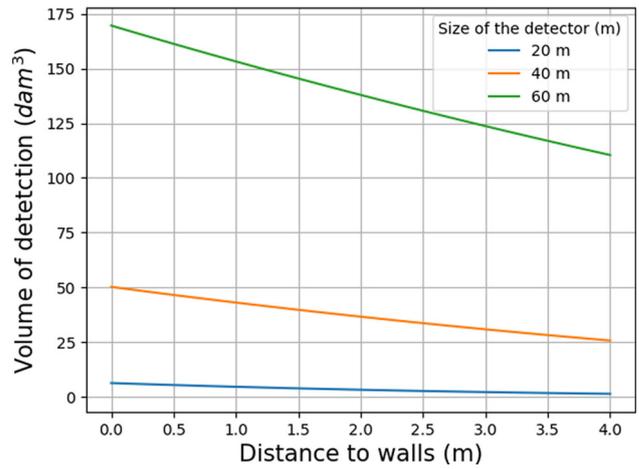

**Fig. 19** Volume of detection as a function of the distance to the walls for the different sizes of detector considered in the previous simulations

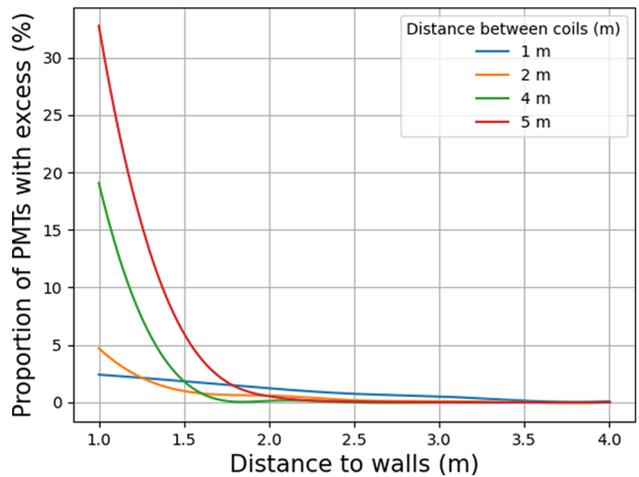

**Fig. 20** Proportion of PMTs with $\Delta B_{perp} > 100$ mG as a function of the distance of the PMTs to the coils for different coil spacings in a detector of 40 m of height and diameter and geomagnetic field of $B = 282.84$ mG and equal components $B_Y = B_Z = 200$ mG

the compensation, especially the larger the size of the detector, as can be seen when $d_W = 1$ m. This can lead to the investment of large amounts of money and decreased efficiency in a considerable number of PMTs. In view of the results obtained in this section, it is proposed to set, if possible, the distance between PMTs and the detector walls to $d_W = 2$ m. In this way, care is taken not to lose too much detection volume and, at the same time, the geomagnetic field compensation is reasonably straightforward, especially with the modifications to the basic coil system that will be discussed later in this manuscript.

3.3 Influence of geomagnetic field

It is also necessary to confirm that the relationship between the ratio of PMTs with excess magnetic field, distance between coils and between PMTs and detector walls analysed from the first simulation holds even for other more extreme values of the Earth's magnetic field. To verify this, several simulations have been performed with different values of the geomagnetic field. For clarity, it is useful to keep in mind the reference frame in Fig. 1, where a generic geomagnetic field vector is shown with respect to the reference frame considered in the detector. Figures 20 and 21 show the proportion of PMTs with excess residual perpendicular magnetic field as a function of the distance between PMTs and the inner surface of the detector, for different values of the distance between coils. In Fig. 20 the simulation is made with a geomagnetic field of $B = 282.84$ mG and equal components $B_Y = B_Z = 200$ mG while in Fig. 21 a terrestrial magnetic field of $B = 636.40$ mG and components $B_Y = B_Z = 450$ mG is considered.

The general behaviour of the ratio of PMTs subjected to excess perpendicular magnetic field as a function of the distance to walls and the distance between coils remains the same as that observed in Fig. 14 with an intermediate value of the Earth's magnetic field. The ratio decreases as the distance to the walls increases and the greater the distance to the walls, the greater the distance between coils, which gives optimum results.

A remarkable, as well as intuitive, fact is that the lower the value of the Earth's magnetic field, the easier it is to compensate for it by this basic coil configuration. In the Fig. 20 plot corresponding to a geomagnetic field of $B = 282.84$ mG it is possible to reduce the proportion of PMTs experiencing excess magnetic field perpendicular to zero by setting a distance to walls greater than or equal to $d_C = 2$ m for any of the depicted values of the distance between coils. As seen in the graph 14, for an intermediate value of the magnetic field of $B = 475$ mG, this total compensation is not achieved for a distance between coils of $d_C = 4$ m or $d_C = 5$ m until reaching a value of the distance to walls of $d_W = 4$ m and for a smaller value of the distance between coils, the distance to walls has to be greater than $d_W = 4$ m. Furthermore, increasing the value of the geomagnetic field, up to $B = 636.4$ mG, yields the results shown in the plot in Fig. 21, where the proportion of PMTs affected is far from zero for the range of wall distances considered. It follows that the lower the latitude at which the detector is located, i.e. the lower the value of the geomagnetic field at its position, the easier it is to compensate for it.

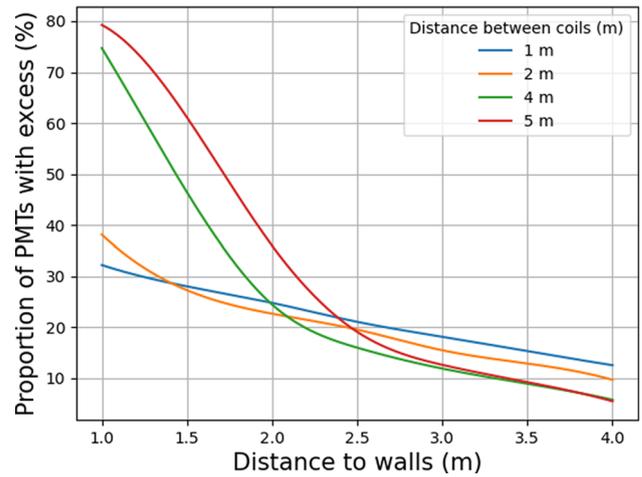

**Fig. 21** Proportion of PMTs with $\Delta B_{perp} > 100$ mG as a function of the distance of the PMTs to the coils for different coil spacings in a detector of 40 m of height and diameter and geomagnetic field of $B = 636.4$ mG and equal components $B_Y = B_Z = 450$ mG

The optimal spacing varies as a function of the distance between the PMTs and the detector walls, just as was observed in the results shown in the graphic in Fig. 14 with a geomagnetic field of intermediate value. The best results are obtained for the largest distances. However, the geomagnetic field does affect to some extent the distance to walls range values over which a given value of the distance between coils is optimal.

For a low value of the earth magnetic field, $B = 282.84$ mG, a distance between coils of $d_C = 1$ m is optimal when the distance to the walls is less than $d_W = 1.2$ m. On the other hand, if $d_W \in [1.2, 1.6]$ m, the distance between coils that gives the best results is $d_C = 2$ m. A distance of $d_C = 4$ m is the best choice for $d_W \in [1.6, 2.1]$ m and, finally, for $d_W > 2.1$ m, $d_C = 5$ m would be the best choice, although for such values of distance to the detector surface the ratio would cancel out at any of the inter-coil distances considered.

On the other hand, with a high value of the geomagnetic field, $B = 636.4$ mG, the ranges of distance to the walls in which each coil distance is optimal varies with respect to the previous case. An inter-coil distance of $d_C = 1$ m is now optimal for $d_W < 1.4$ m, while the appropriate range for $d_C = 2$ m is $d_W \in [1.4, 2.1]$ m, the one associated with $d_C = 4$ m turns out to be $d_W \in [2.1, 3.75]$ m and for $d_C = 5$ m, $d_W > 3.75$ m.

It can be concluded, based on these results and comparing also with those shown in the graph in Fig. 14 corresponding to an intermediate value of the geomagnetic field, that the ideal range of distance to the walls associated to each distance between coils increases its length as the Earth's magnetic field to which the detector is subjected increases, as well as the two extreme values that define the interval also increase.

The effect on the compensation of extreme values of the terrestrial magnetic field has been studied, but only considering the particular case $B_Y = B_Z$. The effect of a magnetic field of intermediate value but whose components have an uneven value is also analysed. Figures 22 and 23 show the proportion of PMTs subjected to excess residual perpendicular magnetic field as a function of their distance from the wall, for various values of the distance between coils in a detector of 40 m of height and diameter and an intermediate value of the geomagnetic field of $B = 475$ mG. In Fig. 22, $B_Y = 250$ mG and $B_Z = 403.88$ mG are considered, while in Fig. 23 it is the other way around, i.e., $B_Y = 403.88$ mG and $B_Z = 250$ mG.

The proportion of PMTs subjected to excess of magnetic field decreases with increasing distance to the walls, regardless of the value of the different magnetic field components. As far as the relationship with the distance between coils is concerned, the behaviour is the same as found so far. The optimal inter-coil distance also increases with the distance to the walls. However, it is observed that the proportion of PMTs affected is more easily minimized when $B_Y > B_Z$, that is, for low latitudes, for a total magnetic field of $B = 475$ mG for a given value of the inter-coil distance, except for $d_C = 5$ m. For a distance between coils with this value, better results are obtained if $B_Y < B_Z$ is set for any value of the distance to the walls. Furthermore, given a specific value of the distance to the walls, the distance between coils that gives the best results is greater in the case where $B_Y < B_Z$.

As far as the current intensity is concerned, the general behaviour found so far holds, regardless of the value of the magnetic field components. The current intensity increases with the distance between coils, and more slightly with the distance of the PMTs from the walls. Naturally, the highest values of the applied current intensity correspond to the set of coils that compensates for the highest component of the magnetic field. Thus, in the case where $B_Y < B_Z$ higher current is applied to the circular coils, while if $B_Y > B_Z$ higher current flows through the rectangular coils. It is also notable that, comparing the current applied in the case where $B_Y = 250$ mG (Fig. 22) with the opposite case where $B_Z = 250$ mG (Fig. 23), a higher current has to be applied to the rectangular coils in the first case than to the circular ones in the other in order to compensate for a field component of equal value. The same is true if $B_Z = 403.88$ mG (Fig. 22) and $B_Y = 403.88$ mG (Fig. 23). The rectangular coils are the ones that need a higher value of the current intensity to compensate for a certain value of the geomagnetic field.

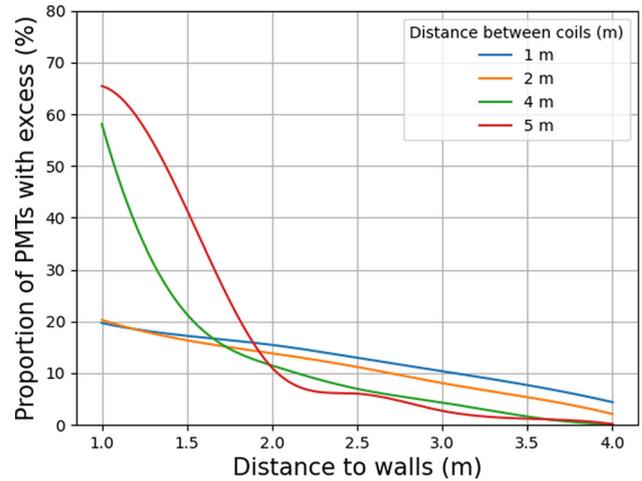

**Fig. 22** Proportion of PMTs with $\Delta B_{perp} > 100$ mG as a function of the distance of the PMTs to the coils for different coil distances in a detector of h = D = 40 m and a value of geomagnetic field of 475 mG with components $B_Y = 250$ mG and $B_Z = 403.88$ mG

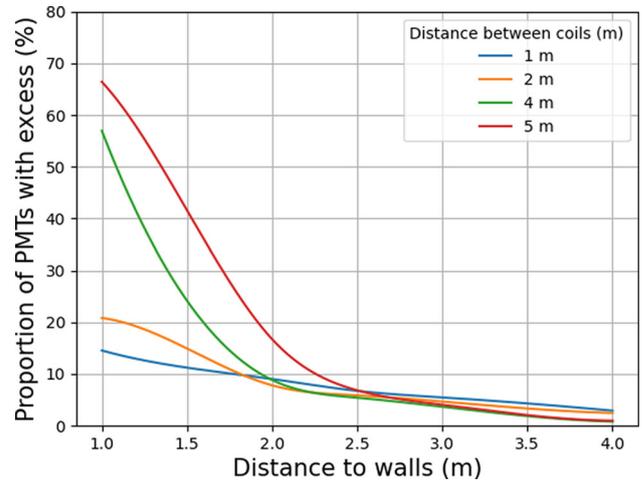

**Fig. 23** Proportion of PMTs with $\Delta B_{perp} > 100$ mG as a function of the distance of the PMTs to the coils for different coil distances in a detector of h = D = 40 m and a value of geomagnetic field of 475 mG with components $B_Y = 403.88$ mG and $B_Z = 250$ mG

3.4 Effect of height to diameter ratio

To conclude the study of the different parameters affecting the geomagnetic field compensation through the configuration of circular and rectangular coils presented in this work, the effect of the detector height to diameter ratio is analysed. This ratio is denoted as R. Figures 24, 25, 26 and 27 show the proportion of PMTs subjected to excess residual perpendicular magnetic field as a function of their distance to the wall, for various values of the distance between coils and a value of the geomagnetic field of $B = 475$ mG in detectors of different values of the ratio $R$. In particular, graphics in Figs. 24 and 25 show the results corresponding to a detector where R = 2 and to a detector with R = 1.5 respectively. In Figs. 26 and 27 the same representation is carried out on wider than tall detectors, in particular for a detector where R = 0.5 and for another where $R = \frac{2}{3}$.

The overall behaviour does not appear to be affected by the detector height-to-diameter ratio. Both if $R > 1$ and if, $R < 1$ the proportion of PMTs subjected to excess magnetic field decreases as the distance from the PMTs to the walls of the detector increases, just as the value of the optimal distance between coils increases with the distance from the PMTs to the walls.

If we compare the results corresponding to Fig. 24 with those of Fig. 25 plot, both with $R > 1$ but different values of this ratio, we find that better results are obtained in Fig. 24 than in Fig. 25, since, for the same value of distance to the walls, the proportion of PMTs subjected to more than 100 mG of perpendicular magnetic field is higher in the latter case. However, this is probably rather due to the fact that the detector considered in Fig. 25 has larger dimensions, something that has already been seen to negatively influence the compensation. The same comparison can be made between the results of Figs. 26 and 27, with $R < 1$ where the same conclusion is reached.

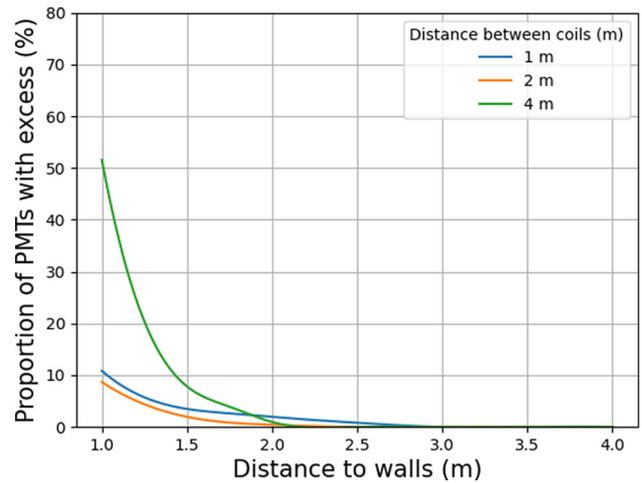

**Fig. 24** Proportion of PMTs with $\Delta B_{perp} > 100$ mG as a function of the distance of the PMTs to the coils for different coil spacings and a value of geomagnetic field of 475 mG in a detector with h = 32 m and D = 16 m (R = 2)

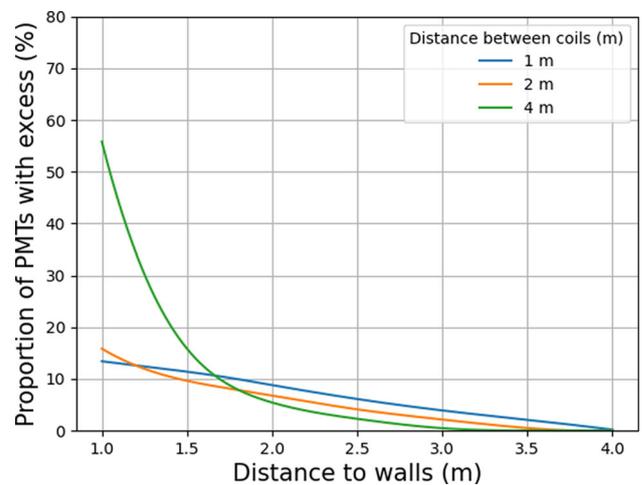

**Fig. 25** Proportion of PMTs with $\Delta B_{perp} > 100$ mG as a function of the distance of the PMTs to the coils for different coil spacings and a value of geomagnetic field of 475 mG in a detector with h = 48 m and D = 32 m (R = 1.5)

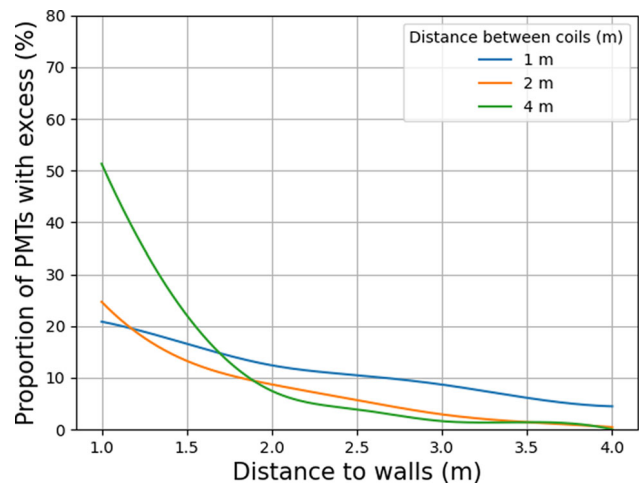

**Fig. 26** Proportion of PMTs with $\Delta B_{perp} > 100$ mG as a function of the distance of the PMTs to the coils for different coil spacings and a value of geomagnetic field of 475 mG in a detector with h = 16 m and D = 32 m (R = 0.5)

A very relevant conclusion that can be obtained comes from comparing the results corresponding to detectors of the same height but different diameter, as well as detectors with the same diameter but different height. In particular, the graphs shown in Figs. 24 and 27 show the results associated to detectors of height h = 32 m, but of different diameter. It can be seen that the plot in Fig. 27 provides worse results, i.e. higher values of the proportion of PMTs subjected to excess magnetic field for the same value of the distance to the walls. Thus, it can be concluded that with the proposed system of coils, for a detector of a certain height, the larger the diameter of the detector, the worse the compensation of the geomagnetic field is.

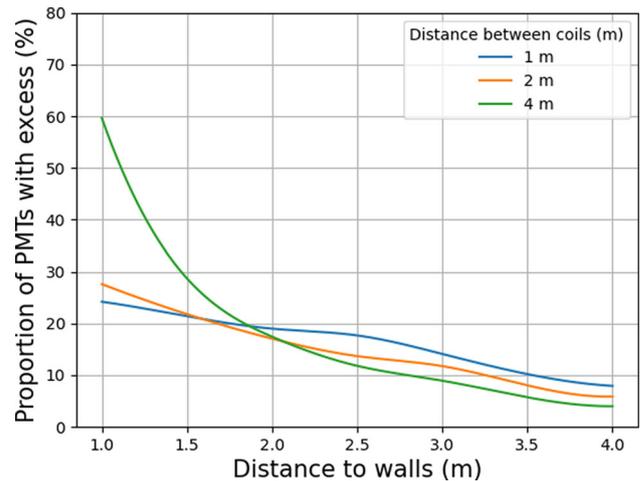

**Fig. 27** Proportion of PMTs with $\Delta B_{perp} > 100$ mG as a function of the distance of the PMTs to the coils for different coil spacings and a value of geomagnetic field of 475 mG in a detector with h = 32 m and D = 48 m ($R = \frac{2}{3}$)

Comparing, on the other hand, the results of Figs. 25 and 26, both associated to detectors of diameter $D = 32$ m but of different height, it is seen that in this case, the best results are offered by the simulation involving the detector of greater height, although in this case the contrast is not as remarkable as in the comparison of different values of the diameter. It is then concluded that in a fixed diameter detector, the greater the height, the better results will be obtained in the compensation of the geomagnetic field perpendicular to the PMTs with the considered coil system.

3.5 Elliptical coils

It has been shown that given a cylindrical detector of certain dimensions, it is possible to compensate the magnetic field by means of a system of circular and rectangular coils in such a way as to minimize the number of PMTs subjected to an excess magnetic field which reduces their efficiency, provided that an optimum distance between coils is chosen and the distance of the PMTs to the walls is sufficiently increased.

However, increasing the distance to the walls implies both a smaller volume of material with which the particles can interact and a smaller photosensitive surface. In addition, given the dimensions of the tank, it may not be possible to choose the optimal distance between coils because of the associated costs or installation issues. This is why it is not uncommon that the number of PMTs subjected to excess magnetic field cannot be minimized as much as desired with the basic configuration of coils presented, and that the proportion of PMTs subjected to excess magnetic field remains relatively high. This subsection therefore deals with the improvement of the coil system for compensation by adding certain coils at specific positions.

As an example, Fig. 28 shows the position of the PMTs even with residual perpendicular magnetic field $\Delta B_{perp} > 100$ in a detector of 40 m height and diameter, where a distance to the walls of $d_W = 2$ m is considered. A total magnetic field of intermediate value of $B = 475$ mG with $B_Y = B_Z$ is considered, so, taking as a reference the results of the Fig. 14, a configuration of rectangular and circular coils with distance between coils of $d_C = 4$ m is established for optimal results.

The proportion of remaining PMTs subjected to excess residual perpendicular magnetic field with this basic loop configuration is 10.08%. While this assumes that 90.08% of the PMTs in the detector will not experience a reduction in efficiency due to the geomagnetic field, and can be considered good results, there is still room for further improvement. Figure 28 shows that the PMTs that still experience excess magnetic field are located at the lids and at the top and bottom edges of the walls, while for the PMTs on the walls at intermediate heights the geomagnetic field is perfectly compensated. The additions to be made to the coil system must then be aimed at exactly those areas where there is excess without affecting the rest of the detector.

A simple way to impact the ends of the detector and decrease the number of remaining PMTs with excess magnetic field in these areas is to increase the number of circular coils at both ends. If the coil at the lower end and the coil at the upper end of the detector are replaced by two solenoids with a certain optimum number of turns, the proportion of PMTs with excess residual magnetic field decreases significantly. Figure 29 shows the remaining PMTs with excess magnetic field, having added one more turn with the same intensity of current at each end to the initial configuration considered in the simulation corresponding to Fig. 28.

The proportion of PMTs with excess of magnetic field has decreased from 10.08% shown in the Fig. 28 to 4.31% in the Fig. 29 by simply adding one more turn at each end, which is a significant improvement.

At this point, the compensation of the magnetic field in the remaining excess PMTs is by no means trivial. From Fig. 29 it can be seen that these remaining PMTs are mainly located at the top and bottom edges of the walls and now to a lesser extent at the lids. Addressing these areas in a localized manner is a complicated task, but a way is proposed below that has been found to give quite satisfactory results.

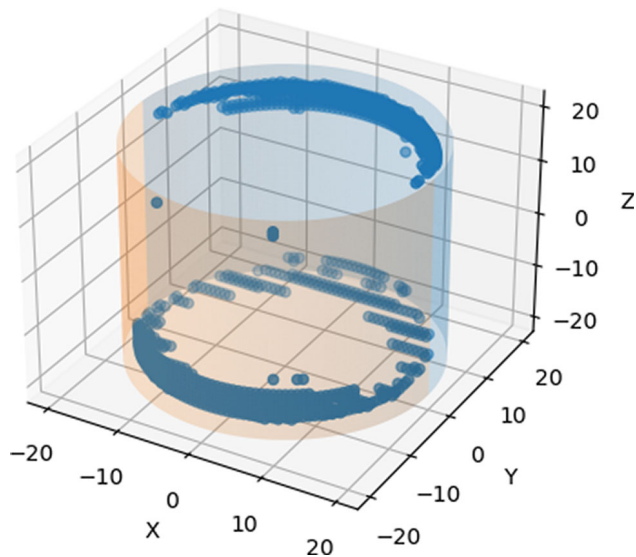

**Fig. 28** Position of PMTs with excess of magnetic field in a detector of $h = D = 40$ m with $d_W = 2$ m to the walls and $B = 475$ mG. A configuration of coils with $d_C = 4$ m between them is chosen

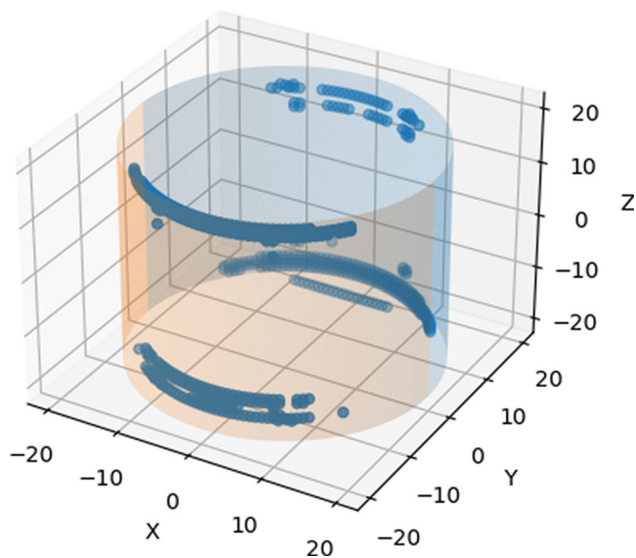

**Fig. 29** Position of PMTs with excess of magnetic field in a detector of $h = D = 40$ m with $d_W = 2$ m to the walls, $d_C = 4$ m between coils and $B = 475$ mG. An extra turn has been added to the top and bottom coils

**Table 3** Position and axes of elliptical coils as a function of the radius of the detector

| Major axis (m) | Minor axis (m) | Center (m) | Lid |
| --- | --- | --- | --- |
| $0.56 \cdot R$ | $0.266 \cdot R$ | $(0, -0.72 \cdot R)$ | Bottom |
| $0.728 \cdot R$ | $0.434 \cdot R$ | $(0, -0.55 \cdot R)$ | Bottom |
| $0.894 \cdot R$ | $0.691 \cdot R$ | $(0, -0.282 \cdot R)$ | Bottom |
| $0.56 \cdot R$ | $0.266 \cdot R$ | $(0, 0.72 \cdot R)$ | Bottom |
| $0.728 \cdot R$ | $0.434 \cdot R$ | $(0, 0.55 \cdot R)$ | Bottom |
| $0.894 \cdot R$ | $0.691 \cdot R$ | $(0, 0.282 \cdot R)$ | Top |

The proposed idea is to place elliptical coils in both bases in very specific positions, so that they mostly affect the areas with excess while affecting the PMTs in the rest of the detector as little as possible. These coils introduce a magnetic field of the order of 10 mG in the three directions $B_x$, $B_y$ and $B_z$ at both ends of the detector. At intermediate heights, the contribution of these coils becomes negligible, decreasing by one or even several orders of magnitude. Therefore, although the addition of these coils may slightly overcompensate the magnetic field perpendicular to some PMTs, the number of PMTs that become subjected to less than 100 mG due to their effect is much higher. The Table 3 shows the dimensions of these coils, as well as the position within the detector where they have to be placed according to their radius. Continuing with the example of the geomagnetic field compensation in a detector of 40 m in height and diameter, the position of the coil would be as shown in the Fig. 30.

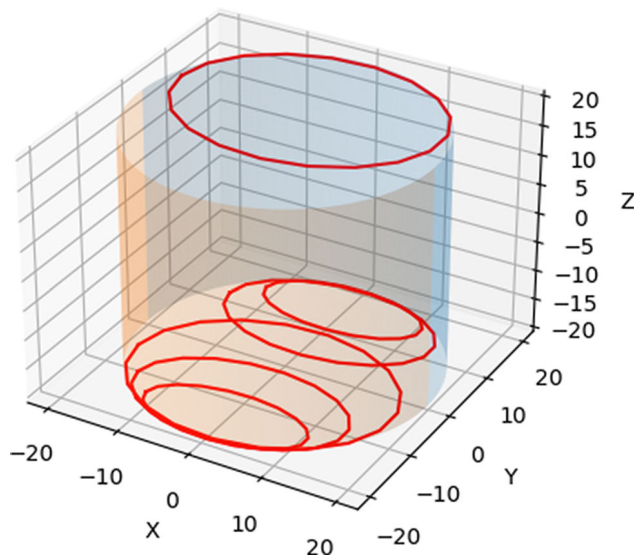

**Fig. 30** Position of the elliptical coils at the detector bases, used to compensate for the excess magnetic field at the edges

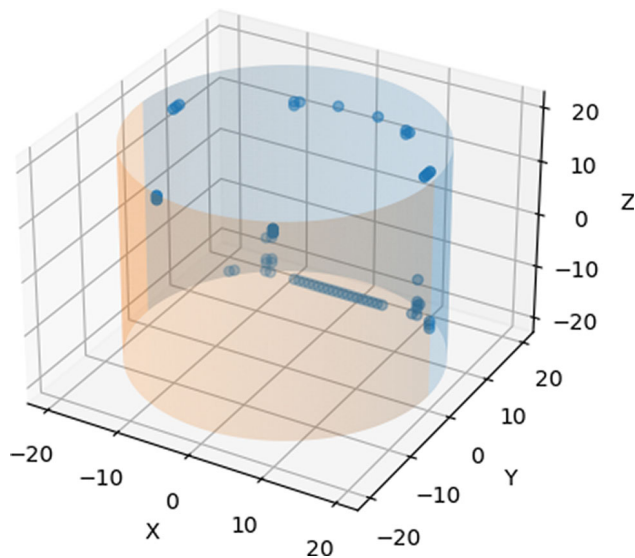

**Fig. 31** Position of PMTs with excess of magnetic field in a detector of $h = D = 40$ m with $d_W = 2$ m to the walls, $d_C = 4$ m between coils and $B = 475$ mG. The basic loop configuration, the additional turn to the end loops, the elliptical base loops and the adjustment of the current intensity in the loops close to the ends are considered

A total of six elliptical coils are used, five in the lower lid and one in the upper lid. It is important to note that, although for the coils located in the lower lid centred at $y < 0$ the direction of the current is such that the field generated compensates the geomagnetic field, as in the rest of the detector's coils, in the other elliptical coils, that is, the coils in the lower lid centred at $y > 0$ and the one in the upper lid, the current flows in the opposite direction and the magnetic field generated has the same direction as the geomagnetic field. This is because at these edges the geomagnetic field is overcompensated, so PMTs in these areas experience more than 100 mG of perpendicular but opposite magnetic field.

To improve the compensation, consideration is also given to adjusting the current of the circular coils near the ends, either by increasing or decreasing the current intensity as necessary. The power supply system will, as a consequence, be slightly more complex, as these particular coils will have to be supplied with a different current intensity than the rest of the circular coils.

Figure 31 finally shows the results of the simulation of the 40 m high and diameter detector, with $d_W = 2$ m distance to the wall and $d_C = 4$ m distance between coils of the basic configuration, whose compensation has been worked so far as an example, after adding the elliptical coils and optimizing their current intensity, as well as after adjusting the current intensity of the second circular coils closest to each of the upper and lower ends.

The proportion of PMTs subjected to excess residual perpendicular magnetic field has been significantly reduced with the addition of the elliptical coils and the adjustment of the current strength of the circular coils near the ends compared to the results of the figure, resulting only from adding one more turn to the upper and lower circular coils. This ratio has decreased from 4.31 to 0.46%.

To summarize, once the optimal basic configuration of circular and rectangular coils has been established, the strategy of minimizing the number of PMTs with excess perpendicular magnetic field, which are located in the covers and at the top and bottom edges of the walls, involves increasing the number of turns of the two circular coils at the ends, adjusting the current intensity of the circular coil closest to it and installing elliptical coils at the edges of both covers. In particular, for the detector example under consideration, these steps have been used to decrease the proportion of PMTs with excess magnetic field from the 10.08% obtained with the basic coil configuration to 0.46%. This means that more than 99.5% of the PMTs in the detector experience an efficiency loss of less than 1% due to the effect of the magnetic fields.

## 4 Conclusions

Experiments in particle physics, such as the indirect search for dark matter candidates, often make use of PMTs (e.g. Cherenkov detectors). The efficiency of this kind of detector is highly affected by the presence of external magnetic fields, and in particular the geomagnetic field. The compensation of the Earth's magnetic field is usually performed applying magnetic shields to the PMTs or adopting methods based on systems of coils.

When compensating the geomagnetic field by means of a coil system in a cylindrical detector using a coil-based system, there are several parameters to be taken into account that affect to a greater or lesser extent the ease of minimizing the magnetic field. There is a clear relationship between the distance between the PMTs and the detector walls and the number of PMTs subjected to excess magnetic field. The larger the distance, the smaller the ratio, eventually reaching zero. However, it should be noted that an increase in the distance to the walls decreases the volume of active material and photosensitive surface, so it is important to take both factors into account when designing the detector.

It is noted that the size of the detector also has an effect on compensation, and that compensation becomes more difficult the larger the dimensions are. In particular, it is observed that, for a certain height, larger diameter has a negative influence, whereas, when the diameter is fixed, more height leads to better results.

The strength of the geomagnetic field, as well as the value of its components, are also key factors to be taken into account. Logically, the higher the geomagnetic field strength, the more difficult it is to compensate using the proposed basic coil configuration, so if the detector is placed at a low latitude, where the geomagnetic field value is lower, the compensation of the geomagnetic field will be easier..

As far as the optimal coil distance is concerned, it depends strongly on the distance to the walls set and more slightly on the other factors mentioned above. Given a certain distance between coils that allows them to be placed symmetrically between the ends of the detector, there is a certain range of distances to the walls in which it results to be the optimal spacing. The greater the distance to the walls, the greater the distance between the coils that yields the best results. However, if the distance between the coils is such that a symmetrical arrangement between the two ends is not possible, the results worsen significantly.

The current intensity applied to both sets of coils is also a crucial parameter, which is affected by the previous ones. It is most dependent on the distance between the coils. The optimum current intensity increases by an average of $\Delta I \approx 35\ A$ per metre increase in the distance between the coils. The optimum current also depends to a lesser extent on the distance from the walls, increasing gradually as the distance from PMTs to the walls increases. The dimensions of the detector also have some influence. The smaller the detector, the smaller the increase of the optimum current intensity with the distance to the walls.

On the other hand, it is also observed that if $B_Y = B_Z$ the optimum current intensity corresponding to the set of rectangular coils is always greater or equal to that of the circular coils. The higher the value of the geomagnetic field, the higher the current that has to be applied to compensate for it, and the higher the current associated with the set of coils that compensates for the component with the highest value.

Taking all the above factors into account, it would be possible in some cases, depending on the dimensions of the detector and the distance from the PMTs to the walls, to fully compensate the geomagnetic field perpendicular to the PMTs with the proposed basic configuration of rectangular and circular coils only. However, since it is not always easy to have a low Earth magnetic field at the detector location or to set the distance between the PMTs and the detector walls sufficiently large, it is most often the case that even with this optimized configuration there are still PMTs with excess magnetic field remaining. If an optimal choice of the above-mentioned parameters has been made, these remaining PMTs have to be located on the lids and on the top and bottom edges of the detector walls, so to minimize the number of these PMTs it is crucial to have a localized impact on these areas. To this end, what has been proposed in this article is to increase the number of turns of the two circular coils at the top and bottom ends, adjust the current intensity of the next two closest coils and place elliptical coils at the edges of both covers with the specific measures that have been established according to the radius of the detector. In the example presented in Sect. 3.5, the proportion of PMTs subject to excess magnetic field has been reduced from the 10.08% achieved with the basic configuration alone, to less than 0.5% by further application of this localized approach.

**Acknowledgements** This work has been developed with financial support from the grant by the SV-PA-21-AYUD/2021/51301 project founded by MICINN and FICYT regional plans. LB acknowledges financial support from the PID2021-125630NB-I00 project funded by MCIN/AEI/10.13039/501100011033/FEDER, UE. FJDC acknowledges financial support from the PID2021-127331NB-I00 project funded by MCIN/AEI/10.13039/501100011033/FEDER, UE.

**Funding** Open Access funding provided thanks to the CRUE-CSIC agreement with Springer Nature.

**Data Availability Statement** The datasets generated during and/or analysed during the current study are available from the corresponding author on reasonable request. This manuscript has associated data in a data repository. [Authors' comment: …].